\documentclass[preprint,12pt,authoryear]{elsarticle}
\usepackage{amssymb,amsmath}
\journal{Journal of Theoretical Biology}

\usepackage{chngcntr}
\usepackage{multirow}
\usepackage{amsmath}

\begin{document}

\begin{frontmatter}

\title{A second-order stability analysis for the continuous model of indirect reciprocity}
\author{Sanghun Lee}
\address{Department of Physics, Pukyong National University, Busan 48513, Korea}
\author{Yohsuke Murase}
\address{RIKEN Center for Computational Science, Kobe, Hyogo 650-0047, Japan}
\author{Seung Ki Baek\corref{cor1}}
\address{Department of Scientific Computing, Pukyong National University, Busan 48513, Korea}
\ead{seungki@pknu.ac.kr}

\begin{abstract}
Reputation is one of key mechanisms to maintain human cooperation, but its
analysis gets complicated if we consider the possibility that reputation does
not reach consensus because of erroneous assessment. The difficulty is
alleviated if we assume that reputation and cooperation do not take binary
values but have continuous spectra so that disagreement over reputation can be
analysed in a perturbative way.
In this work, we carry out the analysis by expanding the dynamics of reputation to the second order of perturbation under the assumption that everyone initially cooperates with good reputation. The second-order theory clarifies the difference between Image Scoring and Simple Standing in that punishment for defection against a well-reputed player should be regarded as good for maintaining cooperation. Moreover, comparison among the leading eight shows that the stabilizing effect of justified punishment weakens if cooperation between two ill-reputed players is regarded as bad. Our analysis thus explains how Simple Standing achieves a high level of stability by permitting justified punishment and also by disregarding irrelevant information in assessing cooperation. This observation suggests which factors affect the stability of a social norm when reputation can be perturbed by noise.
\end{abstract}

\begin{keyword}
Indirect reciprocity \sep Prisoner's dilemma \sep Evolution of cooperation \sep
Perturbation
\end{keyword}

\end{frontmatter}

%\linenumbers
\section{Introduction}

The power and the instability of reputation have attracted interest among
researchers in the field of social evolution~\citep{alexander1987biology}.
Reputation strongly affects our behaviour from early
childhood~\citep{silver2018pint}, but it can turn to a capricious
tyrant: Sometimes a small mistake ruins it without deserving, and it may be
unrecoverable nowadays when digital footprints last forever.
The question is how to make a stable reputation system that recovers from erroneous assessment.

The dynamics of reputation was first analysed in mathematical terms
by considering a norm called Image Scoring~\citep{nowak1998evolution}. It is a
first-order norm in the sense that it assigns reputation to an individual
depending on what she did to her co-player~\citep{nowak2005evolution}.
It also implies that Image Scoring ignores the co-player's reputation.
This is problematic because conditional cooperation crucially hinges on the
ability to base a decision on the co-player's reputation.
By referring to the co-player's reputation,
one must be allowed to refuse to cooperate toward an ill-reputed
co-player without risking her own reputation: Otherwise, well-intentioned
punishment will not be distinguished from malicious defection,
and conditional cooperators cannot thrive in such an environment.
Based on this argument, some authors have advocated `Standing' strategy,
according to which a player loses good reputation only by defecting
against a well-reputed
player~\citep{sugden1986economics,leimar2001evolution}.
`Simple Standing' and its variant called `Consistent Standing' actually belong
to {\it the leading eight}, the set of cooperative norms that are evolutionarily
stable against every behavioural
mutant~\citep{ohtsuki2004should,ohtsuki2006leading}, whereas Image Scoring
does not (Table~\ref{tab:eight}).
The lesson of
`Standing'~\citep{sugden1986economics,leimar2001evolution} holds true
for every member norm of the leading
eight~\citep{ohtsuki2004should,ohtsuki2006leading}:
A well-reputed player's defection against an
ill-reputed co-player should be regarded as good so as to secure cooperation at
the societal level.
We believe that this property should generally be true even
beyond the binary-reputation system~\citep{murase2022social}.

\begin{table}
\caption{Leading eight and Image Scoring (IS). We denote cooperation and defection as $C$ and $D$,
respectively, and a player's reputation as either good ($1$) or bad ($0$).
By $\alpha_{uXv}$, therefore, we mean the reputation assigned to a player who
had reputation $u$ and did $X \in \{C, D\}$ to another player with reputation $v$.
The behavioural rule $\beta_{uv}$ prescribes an action between $C$ and $D$
when the focal player has reputation $u$ and the co-player has reputation $v$.
}
\begin{tabular}{c|cccccccc|cccc}\hline
& $\alpha_{1C1}$ & $\alpha_{1D1}$ & $\alpha_{1C0}$ & $\alpha_{1D0}$ &
 $\alpha_{0C1}$ & $\alpha_{0D1}$ & $\alpha_{0C0}$ & $\alpha_{0D0}$ &
 $\beta_{11}$ & $\beta_{10}$ & $\beta_{01}$ & $\beta_{00}$\\\hline\hline
L1 & 1 & 0 & 1 & 1 & 1 & 0 & 1 & 0 & $C$ & $D$ & $C$ & $C$\\
L2 & 1 & 0 & 0 & 1 & 1 & 0 & 1 & 0 & $C$ & $D$ & $C$ &
$C$\\
L3 & 1 & 0 & 1 & 1 & 1 & 0 & 1 & 1 & $C$ & $D$ & $C$ & $D$\\
L4 & 1 & 0 & 1 & 1 & 1 & 0 & 0 & 1 & $C$ & $D$ & $C$ & $D$\\
L5 & 1 & 0 & 0 & 1 & 1 & 0 & 1 & 1 & $C$ & $D$ & $C$ & $D$\\
L6 & 1 & 0 & 0 & 1 & 1 & 0 & 0 & 1 & $C$ & $D$ & $C$ & $D$\\
L7 & 1 & 0 & 1 & 1 & 1 & 0 & 0 & 0 & $C$ & $D$ & $C$ & $D$\\
L8 & 1 & 0 & 0 & 1 & 1 & 0 & 0 & 0 & $C$ & $D$ & $C$ & $D$\\
IS & 1 & 0 & 1 & 0 & 1 & 0 & 1 & 0 & $C$ & $D$ & $C$ & $D$\\\hline
\end{tabular}
\label{tab:eight}
\end{table}

How to ensure stable cooperation in the presence of error and private reputation rules is still under active
investigation~\citep{uchida2010effect,uchida2013effect,olejarz2015indirect,okada2017tolerant}.
The problem can be illustrated in the following
way~\citep{hilbe2018indirect}:
Suppose that Alice and Bob hold different views on Charlie because
of perception error or their own private assessment rules. Their different views
may also lead to different opinions about what to do to Charlie, especially
if they have adopted strict social norms. Therefore, when David chooses to (or
not to) help Charlie, Alice and Bob will also judge David's action differently.
We can imagine that the disagreement spreads further as time goes by,
and the process may end up with complete segregation of the
society~\citep{oishi2021balanced}.
The crisis can be mitigated if we have an institutional observer who assesses
each member of the society and broadcasts it to all
others~\citep{okada2018solution,radzvilavicius2021adherence}
or if players have tendency to conform to others' average
opinions~\citep{krellner2022pleasing}.
Another way suggested to suppress the chain reaction is to introduce
insensitivity deliberately, e.g., by keeping the existing views about David
unchanged regardless of his action as well as Charlie's
reputation~\citep{quan2019withhold,okada2020two,quan2022keeping}.

Our goal is to understand where the problem arises in general terms.
In our previous work~\citep{lee2021local}, we proposed to regard reputation and
cooperation as continuous variables to calculate the effects of different assessments
in a perturbative way.
This continuum approach offered a simple understanding of the reasons
why some of the leading eight are vulnerable to error in reputation.
In addition, by assuming small difference between the resident and
mutant norms, we derived a threshold for the benefit-to-cost ratio of
cooperation to suppress mutants~\citep{lee2021local}, replacing an earlier prediction relating the
threshold to the probability of
observation~\citep{nowak1998evolution,nowak2005evolution,nowak2006five}.
However, by taking into account only linear-order perturbation,
our previous work failed to address the difference between
first- and second-order norms because the simultaneous action of defection and
bad reputation appears as a second-order effect if
everyone initially cooperates with good reputation.

In this work, we wish to fill this gap by extending the
stability analysis to the second order.
Our analysis identifies second-order effects
such as justified punishment on the stability of the reputation system when perturbation is caused by erroneous assessment.
The analysis in this work differs from exhaustive invasion
analysis~\citep{perret2021evolution} in that we focus on the stability of a pure
system where everyone uses the same norm, rather than evolutionary stability
against mutant norms.

\section{Analysis}
%\subsection{Model}
The basic dynamical process goes as follows:
At every time step, we pick up a
random pair of players, say, $i$ and $j$, as a donor and a recipient,
respectively, from a large population of size $N \gg 1$.
The conventional setting is that they play the one-shot
donation game so that the donor makes a binary decision on
whether to donate $b$ to the recipient by paying $c$,
where $b$ and $c$ represent the benefit and the cost of cooperation, respectively,
with $b>c>0$.
In our continuous version of the donation game, the
donor $i$ chooses the level of donation, $\beta_i \in [0,1]$, so that $b
\beta_i$ is donated to $j$ at the cost of $c \beta_i$. Cooperation and defection
thus correspond to $\beta_i=1$ and $0$, respectively.

Players in the population observe the interaction with probability $q$.
Let $k$ be one of the observers.
We introduce $m_{ki}$ as a continuous variable between $0$ (bad) and $1$ (good), for
describing $i$'s reputation according to $k$'s assessment rule
$\alpha_k = \alpha_k \left( m_{ki}, \beta_i, m_{kj} \right)$.
Likewise, the donation level $\beta_i$ depends on how the donor
assesses herself as well as the recipient, i.e.,
$\beta_i = \beta_i (m_{ii}, m_{ij})$.
As time goes by,
the new assessment replaces the older one, so the averaged dynamics
of $m_{ki}$ in the continuous-time limit can be written as
follows~\citep{hilbe2018indirect,lee2021local}:
\begin{equation}
\frac{d}{dt} m_{ki} = -qm_{ki} + \frac{q}{N-1} \sum_{j \neq i}
\alpha_k \left[ m_{ki}, \beta_i (m_{ii}, m_{ij}), m_{kj} \right],
\label{eq:dynamics}
\end{equation}
where $q$ is the probability of observation.
We assume that
implementation error and perception error occur at a low rate,
so that error plays the role of perturbation to the initial condition without affecting the governing equation itself.

%\subsection{Recovery from disagreement due to error}

Let us assume that the society has
adopted a common social norm $(\alpha, \beta)$, one of whose stationary
fixed points is a
fully cooperative initial state with $m_{ij}=1$ for every pair of $i$ and $j$.
In other words, we consider norms that satisfy
\begin{equation}
\alpha(1,1,1) = \beta(1,1) = 1,
\label{eq:fixed}
\end{equation}
which implies that the norm under consideration maintains a fully
cooperative state unless error occurs.
Error perturbs the players' reputations from this fixed point,
and we are interested in how small
perturbation $\epsilon_{ki} \equiv 1-m_{ki}$ grows over time.
By expanding Eq.~\eqref{eq:dynamics} to the first order in $\epsilon_{ki}$'s, we obtain the following equation:
\begin{equation}
\frac{d}{dt} \epsilon_{ki} \approx
-q(1-A_x) \epsilon_{ki} + qA_y B_x \epsilon_{ii} +
\frac{q}{N-1}
\sum_{j \neq i} [A_y B_y \epsilon_{ij} + A_z \epsilon_{kj}],
\label{eq:disagree}
\end{equation}
where
\begin{subequations}
\begin{align}
A_x &\equiv \left.\partial_x \alpha(x,y,z) \right|_{(1,1,1)}\\
A_y &\equiv \left.\partial_y \alpha(x,y,z) \right|_{(1,1,1)}\\
A_z &\equiv \left.\partial_z \alpha(x,y,z) \right|_{(1,1,1)}\\
B_x &\equiv \left.\partial_x \beta(x,y) \right|_{(1,1)}\\
B_y &\equiv \left.\partial_y \beta(x,y) \right|_{(1,1)}.
\end{align}
\label{eq:derivatives}
\end{subequations}
Each partial derivative can be interpreted as a sensitivity measure: For
example, $A_z$ means how much the judgment about a well-reputed donor's
cooperation toward a recipient is affected when the recipient has
low reputation. Likewise, $B_x$ means how much a donor should
decrease the level of cooperation toward a well-reputed recipient when the
donor's own reputation is low.

To make the story more concrete, we can construct a continuous version of
Simple Standing~\citep{sugden1986economics}, denoted by L3 in Table~\ref{tab:eight},
by using the bilinear and trilinear interpolation methods as
follows:
\begin{subequations}
\begin{align}
\alpha_\text{SS}(x,y,z) &= yz - z + 1\\
\beta_\text{SS}(x,y) &= y.
\end{align}
\label{eq:ss0}
\end{subequations}
Table~\ref{tab:cont} shows the full list of interpolated expressions for
the leading eight and Image Scoring.
If we look at Table~\ref{tab:first}, all the leading eight have $A_x=B_x=0$ and $A_y=B_y=1$ in this linear description.
The only difference among the leading eight lies in $A_z$: That is, L1, L3, L4, and L7 have $A_z=0$, whereas L2, L5, L6, and L8 have $A_z=1$.
As for Image Scoring, the expression is even simpler:
\begin{equation}
\alpha_\text{IS} = \beta_\text{IS} = y.
\label{eq:scoring}
\end{equation}
Note that the linear-order description for Simple Standing and Image Scoring
is given as
\begin{equation}
(A_x, A_y, A_z, B_x, B_y) = (0, 1, 0, 0, 1)
\label{eq:pure}
\end{equation}
in common.
The above values of partial derivatives are interpreted as follows:
Because $A_y=1$, an observer must be
sensitive to the reduction of cooperation toward a well-reputed recipient when
the donor is also well-reputed. A well-reputed donor must reduce the
level of cooperation when he or she meets an ill-reputed recipient
($B_y=1$). At the same time, the norms with Eq.~\eqref{eq:pure}
are indifferent to the donor's reputation ($A_x=B_x=0$) as well as to the
recipient's in assessing the donor's cooperation ($A_z=0$).
In plain language,
one can make an apology because of $B_x=0$, according to which the donor has to
help a well-reputed player even if his or her own reputation is not good.
Then, forgiveness due to $A_x=0$ comes into play because helping a well-reputed
recipient is still regarded as good, even if the donor has once lost reputation.

The difference between Simple Standing and Image Scoring
is manifested in the second-order description
because $A_{yz}=1$ for Simple Standing, whereas $A_{yz}=0$ for Image Scoring as shown in Table~\ref{tab:second},
where $A_{\mu \nu} \equiv \left. \partial^2 \alpha / \partial\mu \partial\nu
\right|_{(1,1,1)}$ and $B_{\mu\nu} \equiv \left.
\partial^2 \beta /\partial \mu \partial \nu \right|_{(1,1)}$.
The second derivatives can be interpreted in the same way as above:
This time, two variables can change simultaneously from the initial state,
so $A_{yz}$ would be related to how much an
observer changes the assessment of a well-reputed donor when the donor reduces
the level of cooperation toward a relatively ill-reputed recipient (see
Appendix~\ref{app:derivative} for more details).

\begin{table}
    \caption{Continuous versions of the leading eight and Image Scoring.
    The assessment rule $\alpha(x,y,z)$ is obtained by applying
    the trilinear interpolation to $\alpha_{xyz}$'s in Table~\ref{tab:eight},
    where $C$ and $D$ correspond to $1$ and $0$, respectively.
    Likewise,
    the behavioural rule $\beta(x,y)$ results from the bilinear interpolation applied
    to $\beta_{xy}$'s. We note that L1 has been nicknamed Contrite Tit-for-Tat in the context of direct reciprocity~\citep{sugden1986economics,takeuchi2007mathematics}.}
    \centering
    \begin{tabular}{l|c|c}\hline
        Norm & $\alpha(x,y,z)$ & $\beta(x,y)$ \\\hline\hline
        L1 & $x+y-xy-xz+xyz$ & $-x+xy+1$\\
        L2 (Consistent Standing) & $x+y-2xy-xz+2xyz$ & $-x+xy+1$\\
        L3 (Simple Standing) & $yz - z + 1$ & $y$\\
        L4 & $-y-z+xy+2yz-xyz+1$ & $y$\\
        L5 & $-z-xy+yz+xyz+1$ & $y$\\
        L6 (Stern Judging) & $-y-z+2yz+1$ & $y$\\
        L7 (Staying) & $x-xz+yz$ & $y$\\
        L8 (Judging) & $x-xy-xz+yz+xyz$ & $y$\\
        Image Scoring & $y$ & $y$\\\hline
    \end{tabular}
    \label{tab:cont}
\end{table}

\begin{table}
    \caption{First-order derivatives of the continuous leading eight and Image Scoring at $(x,y,z)=(1,1,1)$. Note that their differences lie only in $A_z$.}
    \centering
    \begin{tabular}{c|ccc|cc}\hline
        Norm & $A_x$ & $A_y$ & $A_z$ & $B_x$ & $B_y$ \\\hline\hline
        L1 & 0 & 1 & 0 & 0 & 1\\
        L2 & 0 & 1 & 1 & 0 & 1\\
        L3 & 0 & 1 & 0 & 0 & 1\\
        L4 & 0 & 1 & 0 & 0 & 1\\
        L5 & 0 & 1 & 1 & 0 & 1\\
        L6 & 0 & 1 & 1 & 0 & 1\\
        L7 & 0 & 1 & 0 & 0 & 1\\
        L8 & 0 & 1 & 1 & 0 & 1\\
        Image Scoring & 0 & 1 & 0 & 0 & 1\\\hline
    \end{tabular}
    \label{tab:first}
\end{table}

Equation~\eqref{eq:disagree} can be expressed as a linear-algebraic equation
for an $N^2$-dimensional vector $\vec{V} = (\epsilon_{11}, \ldots, \epsilon_{NN})$.
The $N^2 \times N^2$ matrix acted on $\vec{V}$ has the largest eigenvalue in the following form~\citep{lee2021local}:
\begin{equation}
\Lambda_1 = q \left[ -1 + A_x + A_z + A_y (B_x + B_y) \right],
\label{eq:Q}
\end{equation}
and the corresponding eigenvector is
\begin{equation}
\vec{V}_1 = (1,1,\ldots,1)
\label{eq:v1}
\end{equation}
up to a proportionality constant.
If $\Lambda_1>0$, the fixed point in Eq.~\eqref{eq:fixed} is unstable, which is the case
of L2, L5, L6, and L8 because they have $A_z = 1$.
For the others norms of the leading eight as well as for Image Scoring,
this linear-order analysis leaves stability indeterminate by having
$\Lambda_1=0$. From our viewpoint, the important point is that
Simple Standing is not distinguished
from Image Scoring because they have exactly the same first derivatives
[Eq.~\eqref{eq:pure}].

\begin{table}
    \caption{Second-order derivatives of the continuous leading eight and Image Scoring at $(x,y,z)=(1,1,1)$. Note that their differences lie only in $A_{zx}$, $A_{yz}$, and $B_{xy}$.}
    \centering
    \begin{tabular}{c|cccccc|ccc}\hline
        Norm & $A_{xx}$ & $A_{xy}$ & $A_{zx}$ & $A_{yy}$ & $A_{yz}$ & $A_{zz}$ & $B_{xx}$ & $B_{xy}$ & $B_{yy}$ \\\hline\hline
        L1 & 0 & 0 &  0 & 0 & 1 & 0 & 0 & 1 & 0\\
        L2 & 0 & 0 &  1 & 0 & 2 & 0 & 0 & 1 & 0\\
        L3 & 0 & 0 &  0 & 0 & 1 & 0 & 0 & 0 & 0\\
        L4 & 0 & 0 & -1 & 0 & 1 & 0 & 0 & 0 & 0\\
        L5 & 0 & 0 &  1 & 0 & 2 & 0 & 0 & 0 & 0\\
        L6 & 0 & 0 &  0 & 0 & 2 & 0 & 0 & 0 & 0\\
        L7 & 0 & 0 & -1 & 0 & 1 & 0 & 0 & 0 & 0\\
        L8 & 0 & 0 &  0 & 0 & 2 & 0 & 0 & 0 & 0\\
        Image Scoring & 0 & 0 &  0 & 0 & 0 & 0 & 0 & 0 & 0\\\hline
    \end{tabular}
    \label{tab:second}
\end{table}

To proceed, we take into account second-order terms to write down the following equation (see Appendix~\ref{app:second}):
\begin{eqnarray}
\frac{d\epsilon_{ki}}{dt} &=& -q\epsilon_{ki} - \frac{q}{N-1}\sum_{j\neq i} \left[ -A_x \epsilon_{ki} - A_y (B_x \epsilon_{ii} + B_y \epsilon_{ij}) - A_z \epsilon_{kj} \right]\label{eq:e_ki}\\
&&-\frac{q}{N-1} \sum_{j \neq i} \left[ A_y \left( \frac{1}{2} B_{xx} \epsilon_{ii}^2 + B_{xy} \epsilon_{ii} \epsilon_{ij} + \frac{1}{2} B_{yy} \epsilon_{ij}^2 \right) \right.\nonumber\\
&&+\frac{1}{2} A_{xx} \epsilon_{ki}^2 + \frac{1}{2} A_{yy} (B_x \epsilon_{ii} + B_y \epsilon_{ij})^2 + \frac{1}{2} A_{zz} \epsilon_{kj}^2 \nonumber\\
&&\left. +A_{xy} \epsilon_{ki} (B_x \epsilon_{ii} + B_y \epsilon_{ij}) + A_{yz} (B_x \epsilon_{ii} + B_y \epsilon_{ij}) \epsilon_{kj} + A_{zx} \epsilon_{ki} \epsilon_{kj} \right] + \ldots. \nonumber
\end{eqnarray}

We wish to reduce the $N^2$-dimensional dynamics into a one-dimensional one
along the principal eigenvector $\vec{V}_1 = (\epsilon, \epsilon, \ldots, \epsilon)$ [Eq.~\eqref{eq:v1}].
The shape of the eigenvector reflects the symmetry among players,
according to which everyone should deviate from the initial
state by an equal amount, although marginal differences may exist.
To argue the validity of this approximation,
let us write $\epsilon_{ki} = \epsilon + c_{ki}$, where the difference from
the approximation is assumed to be small, i.e., $\left| c_{ki} \right| \ll \epsilon$. We then decompose Eq.~\eqref{eq:e_ki} into two parts, i.e., one for $\epsilon$ and the other for $c_{ki}$. The former is written as
\begin{eqnarray}
\frac{d\epsilon}{dt} &\approx& \Lambda_1 \epsilon
- q\left[ \frac{1}{2}A_{xx} + A_{zx} + \frac{1}{2} A_{zz}
+ \left( A_{xy} + A_{yz} \right) \left( B_x + B_y \right) \right.\nonumber\\
&&+ \frac{1}{2} A_{yy} \left( B_x + B_y \right)^2
\left.+ A_y \left( \frac{1}{2} B_{xx} + B_{xy} + \frac{1}{2} B_{yy} \right) \right] \epsilon^2,
\label{eq:de}
\end{eqnarray}
and the latter obeys the following dynamics:
\begin{eqnarray}
\frac{dc_{ki}}{dt} &\approx& q\{(-1+A_x)c_{ki}+A_y B_x c_{ii}\}+\frac{q}{N-1}\sum_{j \neq i}(A_y B_y c_{ij}+A_z c_{kj})\nonumber\\
&&-\frac{q\epsilon}{N-1}\sum_{j \neq i}[A_y\{B_{xx}c_{ii}+B_{xy}(c_{ii}+c_{ij})+B_{yy}c_{ij}\}\nonumber\\
&&+A_{xx}c_{ki}+A_{yy}(B_x c_{ii}+B_y c_{ij})(B_x+B_y)+A_{zz}c_{kj}\nonumber\\
&&+(A_{xy}+A_{yz})(B_x c_{ii}+B_y c_{ij})+(A_{xy}c_{ki}+A_{yz}c_{kj})(B_x+B_y)\nonumber\\
&&+A_{zx}(c_{ki}+c_{kj})],
\label{eq:dcki}
\end{eqnarray}
where $\epsilon$ is regarded as a constant.
For the norms with $\Lambda_1=0$, i.e., L1, L3, L4, L7, and Image Scoring,
the largest eigenvalue of Eq.~\eqref{eq:dcki}
is either negative or zero (see Appendix~\ref{app:dcki}). Specifically, it is $-2\epsilon q$ for L1 and L3 (Simple Standing), and zero for L4, L7, and Image Scoring. The point is that $c_{ki}$ does not grow over time, so that the one-dimensional dynamics in Eq.~\eqref{eq:de} remains valid when we consider the leading eight and Image Scoring.
If we plug the partial derivatives in Tables~\ref{tab:first} and \ref{tab:second} into Eq.~\eqref{eq:de}, the dynamics of $\epsilon$ reduces to
\begin{equation}
    \frac{d\epsilon}{dt} \approx \Lambda_1 \epsilon - q (A_{zx} + A_{yz} +
    B_{xy}) \epsilon^2.
    \label{eq:competition}
\end{equation}
For Simple Standing, we arrive at
\begin{equation}
\frac{d\epsilon}{dt} \approx -q\epsilon^2,
\label{eq:1/t}
\end{equation}
which admits a solution in the form of $\epsilon \sim 1/t$.
Its diverging time scale is self-consistent with our assumption that $\epsilon$ can be approximated as a constant in Eq.~\eqref{eq:dcki}.
As for Image Scoring, on the other hand,
$d\epsilon / dt$ is zero up to the second order of perturbation,
meaning that the restoring force toward the initial cooperation is still absent.
Among the nine norms under consideration,
L1 is predicted to show the fastest recovery with the aid of $B_{xy}=1$. Its nonvanishing $B_{xy}$ originates from $\beta_{00}=C$ in
Table~\ref{tab:eight}: By prescribing cooperation between two ill-reputed
players, it promotes recovery more efficiently than Simple Standing, although it
again has a diverging time scale.
Table~\ref{tab:stability} shows the results of recovery analysis
when applied to each of the leading eight and Image Scoring.
These results are consistent with the numerical simulation as shown in the next section.

\begin{table}
\centering
    \caption{
    Stability of the initial cooperative
    state with $\epsilon=0$ under each of the leading eight
    and Image Scoring, according to
    our second-order analysis of the continuous model.
    }
    \begin{tabular}{l|l|c}\hline
    Norm & Dynamics & Stability of $\epsilon=0$\\\hline\hline
    L1 & $d\epsilon/dt = -2q\epsilon^2 + \ldots$ & Stable \\
    L2 (Consistent Standing) & $d\epsilon/dt = q\epsilon + \ldots$ & Unstable\\
    L3 (Simple Standing) & $d\epsilon/dt = -q\epsilon^2 + \ldots$ & Stable\\
    L4 & $d\epsilon/dt = 0 + \ldots$ & Neutral\\
    L5 & $d\epsilon/dt = q\epsilon + \ldots$ & Unstable\\
    L6 (Stern Judging) & $d\epsilon/dt = q\epsilon + \ldots$ & Unstable\\
    L7 (Staying) & $d\epsilon/dt = 0 + \ldots$ & Neutral\\
    L8 (Judging) & $d\epsilon/dt = q\epsilon + \ldots$ & Unstable\\
    Image Scoring & $d\epsilon/dt = 0 + \ldots$ & Neutral\\\hline
    \end{tabular}
    \label{tab:stability}
\end{table}

\section{Results}

\begin{figure}
    \centering
    \includegraphics[width = 0.8\textwidth]{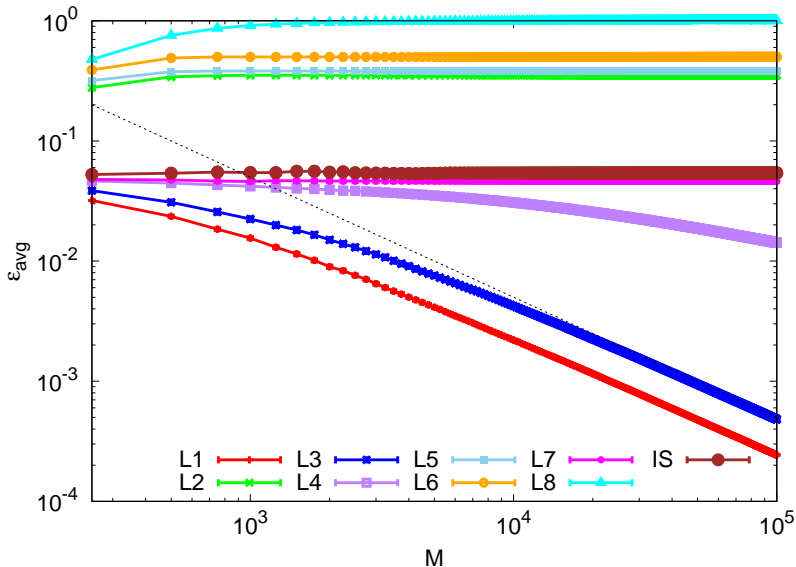}
    \caption{Recovery from disagreement under the leading eight and Image
    Scoring (abbreviated as IS). Consistently with Table~\ref{tab:stability},
    the average of $\epsilon_{ki}$ converges to zero only under L1
    and L3,
    whereas it remains finite under L7 and IS. Differently from the prediction
    in Table~\ref{tab:stability}, L4 shows an extremely slow decay of
    $\epsilon_\text{avg}$, which may be attributed to its high-order terms.
    The other four norms worsen
    a small decline in reputation as seen from the gradual increase of
    $\epsilon_\text{avg}$, consistently with $\Lambda_1>0$.
    The black dotted line shows the decay inversely proportional to time for
    comparison.
    The population size is $N=50$, and the
    curves are average results over $10^2$ samples. The error bars are smaller
    than the symbol size.
    }
    \label{fig:recovery}
\end{figure}

To check the recovery process from error, we conduct numerical simulation.
The code is identical to the one used in our previous work~\citep{lee2021local}:
Let us assume that every player uses the same $\alpha$ and $\beta$.
We consider a population of size $N \gg 1$ by using an $N \times N$ image matrix, $\{ m_{ij} \}$.
The matrix elements are random numbers uniformly drawn from $[0.9,1.0]$.
At each time
step, we pick up a random pair of players $i$ and $j$, the former as the donor
and the latter as the recipient.
The donor chooses the donation level $\beta_i (m_{ii}, m_{ij})$.
This interaction is observed by each of the other members in the
society with probability $q$.
Each observer, say, $k$, updates $m_{ki}$ according to
$\alpha_k \left[ m_{ki}, \beta_i(m_{ii}, m_{ij}), m_{kj} \right]$.
We repeat the above procedure $M$ times, so $M$ can be regarded as a time index.

Our Monte Carlo results in Fig.~\ref{fig:recovery} corroborates the predictions
in Table~\ref{tab:stability}. The instability of L2, L5, L6, and L8 is already
clear from the fact that $\Lambda_1 > 0$. The stability of L1 and L3 is,
however, correctly predicted only when we go through the second-order
stability analysis [Eq.~\eqref{eq:de}]. The behaviour of $\epsilon \sim 1/t$ is
also confirmed by this simulation.
Still, the analysis leaves the stability of L4 and L7
undetermined, and Fig.~\ref{fig:recovery} suggests that the average of
$\epsilon_{ij}$'s will converge to a finite value for L7, whereas it decays
extremely slowly in the case of L4. The latter behaviour is not captured by our
present study, and a higher-order analysis will be required to
understand it.
Under Image Scoring, the initial state certainly has neutral stability.
To sum up, our second-order stability analysis successfully explains the
difference between Simple Standing and Image Scoring: The former recovers from
erroneous disagreement, albeit slowly, whereas the latter does not.

\begin{figure}
    \centering
    \includegraphics[width = 0.8\textwidth]{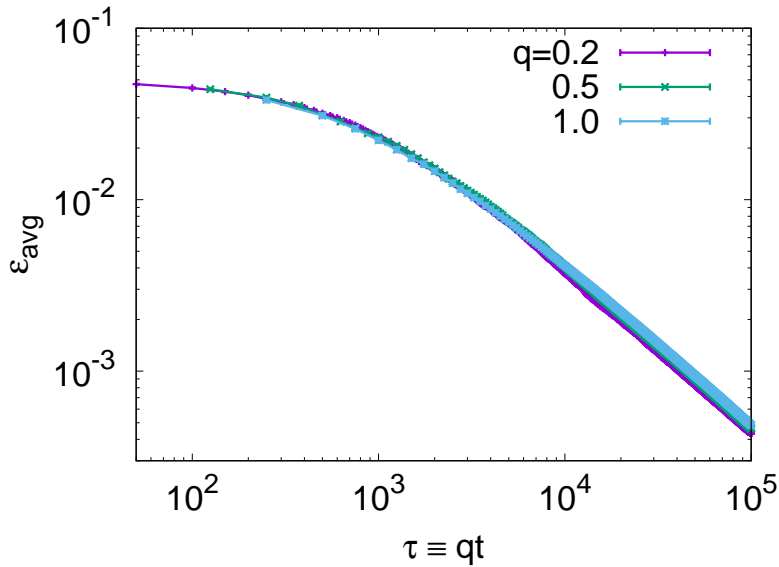}
    \caption{Effects of the observation
    probability $q$ on the recovery dynamics of Simple Standing (L3).
    The horizontal axis represents a rescaled time variable $\tau \equiv qt$,
    where $t$ means time steps.
    The other simulation parameters are the same as in Fig.~\ref{fig:recovery}.
    }
    \label{fig:qt}
\end{figure}

The governing equation [Eq.~\eqref{eq:dynamics}] suggests that the observation
probability $q$ will only change the overall time scale. That is, if we rescale
time by defining $\tau \equiv qt$, the derivative with respect to $\tau$
will become independent of $q$. Figure~\ref{fig:qt}
indeed shows that the the average deviations from the initial state,
$\epsilon_{\rm avg}$, as functions of $\tau$ behave similarly
regardless of $q$ in the case of Simple Standing.
Another point to mention is that the cost-benefit ratio is irrelevant to this
stability analysis because it does not enter the dynamics
[Eq.~\eqref{eq:dynamics}]. The ratio becomes
crucial when the symmetry among players breaks down, e.g., when some players
adopt a different norm from the existing one. In such a case, the cost-benefit
ratio will determine whether the mutant norm can invade the
population~\citep{lee2021local}.

\section{Discussion and Summary}
\label{sec:discuss}

The continuous dynamics of reputation and behaviour opens up the possibility to
apply powerful analytic tools to the study of indirect reciprocity.
For this reason,
the continuum framework is one of the most convenient ways to study general
conditions for norms to be successful when variations in $\alpha$ and $\beta$
can be treated in a perturbative way.
This idea is especially relevant when norms resist abrupt changes,
as suggested by empirical
observations~\citep{mackie2015social,amato2018dynamics}.
A comprehensive and systematic
investigation of this framework
would thus greatly enhance our understanding of cooperation
through indirect reciprocity.
However, our previous linear-order solution provided an inconclusive or
incorrect answer
when applied to the well-known first-order norm called Image Scoring because
the solution did not consider the second-order effects such as justified punishment.
In addition, the difference among the leading eight in their responses to error,
despite the ability of justified punishment shared by all of them, remained
unanswered there.

In this work, we have shown how one can go beyond the linear-order analysis
and identified where the difference arises, which will be an important
piece of information in analysing a social norm.
We have expanded the governing equation [Eq.~\eqref{eq:dynamics}] to the second order of $\epsilon_{ki}$'s and argued the reason that the effective dynamics can reduce to a one-dimensional one. Such reduced dynamics agrees well with numerical simulation for each of the leading eight as well as for Image Scoring.
The success of this one-dimensional reduction must be related to the mean-field
nature of Eq.~\eqref{eq:dynamics}, although it involves intricate three-body
interaction among a donor, a recipient, and an observer at each time step.
The mean-field assumption can readily be justified in a well-mixed
population, where everyone participates in the interaction with equal
probability at a symmetric position. The effect of heterogeneity among players
on the process of rebuilding a consensus remains as a future work.

We also mention the following limitations of our approach:
First, we have examined only the vicinity of a fixed point at which everyone
cooperates with a good reputation. If the initial state is arbitrarily far away
from the fixed point, the analysis loses validity in principle.
In addition, the second-order stability analysis may fail to address the
difference among high-order norms, as can be seen from the fact that the
behaviour of L4 in Fig.~\ref{fig:recovery} is not described by our analysis.
Another limitation is that our approach considers the presence of error only to
escape from the initial state, leaving the subsequent dynamics
unchanged [Eq.~\eqref{eq:dynamics}].
If error occurs continuously in time with a small rate of $\zeta$,
the dynamics of $\epsilon$ for Simple Standing [Eq.~\eqref{eq:1/t}] will be
rewritten as
\begin{equation}
\frac{d\epsilon}{dt} \approx -q\epsilon^2 + \zeta.
\label{eq:zeta}
\end{equation}
In a stationary state, we would
thus expect $\epsilon_{\rm st} \propto \sqrt{\zeta}$, which is much greater
than the magnitude of $\zeta$ itself. Such amplification of error is due to the
lack of linear stability at $\epsilon=0$. In other words, the recovery process
is so slow that any finite error rate can easily keep the system away from
the state of $\epsilon=0$. In this sense, the initial cooperative state is
only weakly protected from erroneous assessment even under the action of L1 or
L3.

\begin{table}
    \caption{Characteristics of the leading eight for successful recovery from error. The left two columns show mathematical representations in the continuous and binary models, respectively. The third column means which of the leading eight satisfy the condition, and the last column explains how to interpret the characteristics.}
    \centering
    \begin{tabular}{cc|cc}\hline
        Continuous & Binary & Norms & Interpretation\\\hline\hline
        $\left\{\begin{array}{l}\alpha(1,1,1)=1\\ \beta(1,1)=1\end{array}\right.$ & $\left\{\begin{array}{l}\alpha_{1C1}=1 \\ \beta_{11}=C \end{array}\right.$ & L1--L8 & Maintenance of cooperation\\\hline
        $A_x=0$ & $\alpha_{0C1}=1$ & L1--L8 & Forgiveness\\
        $B_x=0$ & $\beta_{01}=C$ & L1--L8 & Apology\\
        $A_y=1$ & $\alpha_{1D1}=0$ & L1--L8 & Identification of defectors\\
        $B_y=1$ & $\beta_{10}=D$ & L1--L8 & Punishment\\
        $A_{yz}>0$ & $\alpha_{1D0}=1$ & L1--L8 & Justification of punishment\\\hline
        $A_z=0$ & $\alpha_{1C0}=1$ & L1,L3,L4,L7 & Approval for cooperation\\
        $A_{zx}=0$ & $\alpha_{0C0}=1$ & L1,L3 & to the ill-reputed\\
        \hline
        \multirow{2}{*}{$B_{xy}=1$} & \multirow{2}{*}{$\beta_{00}=C$}
        & \multirow{2}{*}{L1} & Cooperation between\\
        & & & the ill-reputed\\\hline
    \end{tabular}
    \label{tab:summary}
\end{table}

Let us mention a few points on the leading eight from the viewpoint of our
second-order stability analysis (see Table~\ref{tab:summary}).
In our continuum framework, all of the leading eight have
\begin{equation}
    \begin{aligned}
    A_y=B_y=1\\
    A_x=B_x=0
    \end{aligned}
    \label{eq:common1}
\end{equation}
and these slope values are related to the basic properties for being nice, retaliatory, apologetic, and forgiving~\citep{ohtsuki2006leading}. If we look further into their second derivatives, we find another common feature that
they all have
\begin{equation}
    A_{yz}>0
    \label{eq:common2}
\end{equation}
so as to justify punishment on an ill-reputed player (Appendix~\ref{app:derivative}).
The linear-order stability analysis shows that L2, L5, L6, and L8 are nevertheless unstable because $\Lambda_1 = qA_z > 0$ [see Eq.~\eqref{eq:Q}].
This observation imposes an additional condition that
\begin{equation}
    A_z=0,
    \label{eq:common3}
\end{equation}
which means that a well-reputed player's cooperation to an ill-reputed player should be regarded as good.
Among the leading eight, L1, L3, L4, and L7 share this property, and we note that they actually show the quickest recovery from error in the original discrete version of the model~\citep{hilbe2018indirect}.
Now when it comes to L4 and L7, which lack restoring force in spite of all the above three conditions [Eqs.~\eqref{eq:common1} to \eqref{eq:common3}], it turns out that the effect of justified punishment is exactly cancelled out by $A_{zx}=-1$ [see Eq.~\eqref{eq:competition}].
This second derivative is related with how to judge cooperation between two ill-reputed players (see $\alpha_{0C0}$ in Table~\ref{tab:eight}).
L4 and L7 regard such cooperation as bad ($\alpha_{0C0}=0$) and fail to recover from error. When it is regarded as good ($\alpha_{0C0}=1$), the recovery process succeeds, albeit slowly, as we see from L1 and L3.
Our continuum approach thus suggests that
the following property stabilizes L1 and L3 in the second-order analysis:
\begin{equation}
    A_{zx} = 0.
\end{equation}
Together with Eq.~\eqref{eq:common3}, this last condition implies that cooperation to an ill-reputed player should be regarded as good, irrespective of the donor's reputation.
All these features are shared by L1 and L3 in common, and their sole difference
in assessment lies in $\alpha_{0D0}$ (Table~\ref{tab:eight}).
If we focus on L3 (Simple Standing), when you encounter an ill-reputed player, it is always good for your reputation whether you choose to cooperate or punish the co-player. Only defection against a well-reputed player is regarded as bad ($\alpha_{1D1} = \alpha_{0D1} = 0$).
By disregarding irrelevant information on reputation, Simple Standing achieves a high level of stability in a noisy environment.
Note that invasion analysis also shows that those who
defect against ill-reputed individuals do not have to be regarded as bad for a
social norm to have evolutionary
stability~\citep{perret2021evolution}.
Finally, L1 shows faster recovery than L3 because it prescribes cooperation
between two ill-reputed players by having $\beta_{00}=C$.

One of the open issues that remain untouched in this paper is the evolutionary stability against mutants.
Our previous paper analytically obtained the critical benefit-to-cost ratio above which close mutants are driven out, and we may improve the theory by incorporating the second-order effects properly as is done in this paper.
This is a promising direction to deepen our understanding of the mechanism to sustain cooperation.

\section*{Acknowledgements}
S.K.B. acknowledges support by Basic Science Research Program through the National Research Foundation of Korea (NRF) funded by the Ministry of Education (NRF-2020R1I1A2071670).
Y.M. acknowledges support from Japan Society for the Promotion of Science (JSPS) (JSPS KAKENHI; Grant no. 18H03621 and Grant no. 21K03362).
We appreciate the APCTP for its hospitality during the completion of this work.

\appendix
\counterwithin{figure}{section}
\numberwithin{equation}{section}

\section{Second derivatives}
\label{app:derivative}

The second derivative of $\alpha$ with respect to $y$ and $z$ at the fixed point $(x,y,z)=(1,1,1)$ can be approximated as
\begin{equation}
    A_{yz}\approx \frac{\alpha(1,1,1) - \alpha(1,1-h,1) - \alpha(1,1,1-h) + \alpha(1,1-h,1-h)}{h^2}
\end{equation}
with a small parameter $h$.
To see the meaning of $A_{yz}>0$ clearly, let us consider a special case where $\alpha$ has no $z$-dependence when $x=y=1$. Then, the positivity of $A_{yz}$ is equivalent to
\begin{equation}
    \alpha(1,1-h,1-h) > \alpha(1,1-h,1).
\end{equation}
In other words, when one reduces the level of cooperation ($y=1-h$), it is regarded as good if the co-player has bad reputation ($z=1-h$). We can generally consider the case with $z$-dependence, and the point is that one earns better reputation by punishing an ill-reputed player than not.

Likewise, we can approximate $A_{zx}$ as
\begin{equation}
    A_{zx}\approx \frac{\alpha(1,1,1) - \alpha(1-h,1,1) - \alpha(1,1,1-h) + \alpha(1-h,1,1-h)}{h^2}.
\end{equation}
Again, let us assume that $\alpha$ has no $x$-dependence when $y=z=1$ for convenience of explanation. This assumption is especially relevant to the leading eight because they all have $A_x=0$.
Then, the negativity of $A_{zx}$ means the following:
\begin{equation}
    \alpha(1-h, 1, 1-h) < \alpha(1, 1, 1-h).
    \label{eq:second2}
\end{equation}
Note that the right-hand side is effectively the same as $\alpha(1,1,1)=1$
for L1, L3, L4, and L7 because they have $A_z=0$.
The inequality in Eq.~\eqref{eq:second2} implies that
cooperation ($y=1$) between two ill-reputed players ($x=z=1-h$) is regarded as bad by L4 and L7, for which $A_{zx}=-1$.

By the same token, we can say that $B_{xy}>0$ of L1 under the condition that
$B_x=0$ implies
\begin{equation}
\beta(1-h, 1-h) > \beta(1,1-h),
\end{equation}
which corresponds to $\beta_{00}=C$ and $\beta_{10}=D$ in
Table~\ref{tab:eight}. The latter prescription is required to punish defectors,
but the former is nontrivial: If both the donor and the recipient are
ill-reputed, L1 tells the donor to cooperate, which speeds up the recovery of
the reputation.

\section{Second-order perturbation}
\label{app:second}

When $\epsilon_{ij}$'s are
small parameters, the
second-order perturbation for $\beta$ can be written as follows:
\begin{eqnarray}
&&\beta (m_{11}, m_{1j}) = \beta(1-\epsilon_{11}, 1-\epsilon_{1j})\\
&\approx& 1 - B_x \epsilon_{11} - B_y \epsilon_{1j} +
\frac{1}{2} B_{xx} \epsilon_{11}^2 + B_{xy} \epsilon_{11} \epsilon_{1j}
+ \frac{1}{2} B_{yy} \epsilon_{1j}^2 \\
&\equiv& 1 - \kappa.
\end{eqnarray}
Here, we write $\kappa \equiv \kappa^{(1)} + \kappa^{(2)}$, where
$\kappa^{(1)} \equiv B_x \epsilon_{11} + B_y
\epsilon_{1j}$ and $\kappa^{(2)} \equiv -
\left( \frac{1}{2} B_{xx} \epsilon_{11}^2 + B_{xy} \epsilon_{11}
\epsilon_{1j} + \frac{1}{2} B_{yy} \epsilon_{1j}^2 \right)$ are first- and
second-order corrections, respectively.
The second-order perturbation for $\alpha$ is also straightforward:
\begin{eqnarray}
&&\alpha [m_{1i}, \beta_i (m_{ii}, m_{ij}), m_{1j}]
\approx \alpha(1-\epsilon_{1i}, 1-\kappa, 1-\epsilon_{1j})\\
&\approx& 1 - A_x \epsilon_{1i} - A_y \kappa - A_z
\epsilon_{1j}
+ \frac{1}{2} A_{xx} \epsilon_{1i}^2 + \frac{1}{2} A_{yy}
\left(\kappa^{(1)} \right)^2 + \frac{1}{2} A_{zz}
\epsilon_{1j}^2\nonumber\\
&&+ A_{xy} \epsilon_{1i} \kappa^{(1)}
+ A_{yz} \kappa^{(1)} \epsilon_{1j}
+ A_{zx} \epsilon_{1i} \epsilon_{1j}.
\end{eqnarray}

\section{Eigenvalues of the system of $c_{ki}$'s}
\label{app:dcki}

For $N=2$, the full eigenvalue structure of Eq.~\eqref{eq:dcki} is obtained as follows:
\begin{eqnarray}
\lambda_1 &=& q[-1+A_x-A_z\nonumber\\
&&-\{A_{xx}-A_{zz}+(A_{xy}-A_{yz})(B_x+B_y)\}\epsilon]\nonumber\\
\lambda_2 &=& q[-1+A_x-A_z+A_yB_x-A_yB_y\nonumber\\
&&-\{A_{xx}-A_{zz}+2A_{xy}B_x-2A_{yz}B_y+A_{yy}(B_x-B_y)(B_x+B_y)\nonumber\\
&&+A_y(B_{xx}-B_{yy})\}\epsilon]\nonumber\\
\lambda_3 &=& q[-1+A_x+A_z\nonumber\\
&&-\{A_{xx}+2A_{xz}+A_{zz}+(A_{xy}+A_{yz})(B_x+B_y)\}\epsilon]\nonumber\\
\lambda_4 &=& q[-1+A_x+A_z+A_yB_x+A_yB_y\nonumber\\
&&-\{A_{xx}+2A_{xz}+A_{zz}+2(A_{xy}+A_{yz})(B_x+B_y)\nonumber\\
&&+A_{yy}(B_x+B_y)^2+A_y(B_{xx}+2B_{xy}+B_{yy})\}\epsilon].
\end{eqnarray}
If we consider the leading eight, we can readily obtain eigenvalues for $N=2,\ldots,5$ and generalize the patterns: For example, L1 has the following structure:
\begin{eqnarray}
\lambda_1^{(N^2-2N+1)} &=& \frac{1}{N-1}\{-(N-1)+\epsilon\}q\nonumber\\
\lambda_2^{(N-1)} &=& -\frac{1}{N-1}\{N+(N-4)\epsilon\}q \nonumber\\
\lambda_3^{(N-1)} &=& -(1+\epsilon)q\nonumber\\
\lambda_4^{(1)} &=& -4\epsilon q,
\end{eqnarray}
where the superscript on each eigenvalue indicates its multiplicity.
For L3, we find a similar result:
\begin{eqnarray}
\lambda_1^{(N^2-2N+1)} &=& \frac{1}{N-1}\{-(N-1)+\epsilon\}q\nonumber\\
\lambda_2^{(N-1)} &=& -\frac{1}{N-1}(N-2\epsilon)q\nonumber\\
\lambda_3^{(N-1)} &=& -(1+\epsilon)q\nonumber\\
\lambda_4^{(1)} &=& -2\epsilon q.
\end{eqnarray}
Finally, the structure becomes even simpler for L4 and L7:
\begin{eqnarray}
\lambda_1^{(N^2-N)} &=& -(1-\epsilon)q\nonumber\\
\lambda_2^{(N-1)} &=& -\frac{N}{N-1}(1-\epsilon)q\nonumber\\
\lambda_3^{(1)} &=& 0.
\end{eqnarray}
In every case, the largest eigenvalue is the last one, which is either negative or zero.

%\bibliographystyle{elsarticle-harv}
%\bibliography{newton}

\begin{thebibliography}{27}
\expandafter\ifx\csname natexlab\endcsname\relax\def\natexlab#1{#1}\fi
\providecommand{\url}[1]{\texttt{#1}}
\providecommand{\href}[2]{#2}
\providecommand{\path}[1]{#1}
\providecommand{\DOIprefix}{doi:}
\providecommand{\ArXivprefix}{arXiv:}
\providecommand{\URLprefix}{URL: }
\providecommand{\Pubmedprefix}{pmid:}
\providecommand{\doi}[1]{\href{http://dx.doi.org/#1}{\path{#1}}}
\providecommand{\Pubmed}[1]{\href{pmid:#1}{\path{#1}}}
\providecommand{\bibinfo}[2]{#2}
\ifx\xfnm\relax \def\xfnm[#1]{\unskip,\space#1}\fi
%Type = Book
\bibitem[{Alexander(1987)}]{alexander1987biology}
\bibinfo{author}{Alexander, R.D.}, \bibinfo{year}{1987}.
\newblock \bibinfo{title}{The Biology of Moral Systems}.
\newblock \bibinfo{publisher}{Aldine~de~Gruyter}, \bibinfo{address}{New York}.
%Type = Article
\bibitem[{Amato et~al.(2018)Amato, Lacasa, D{\'\i}az-Guilera and
  Baronchelli}]{amato2018dynamics}
\bibinfo{author}{Amato, R.}, \bibinfo{author}{Lacasa, L.},
  \bibinfo{author}{D{\'\i}az-Guilera, A.}, \bibinfo{author}{Baronchelli, A.},
  \bibinfo{year}{2018}.
\newblock \bibinfo{title}{The dynamics of norm change in the cultural evolution
  of language}.
\newblock \bibinfo{journal}{Proc. Natl. Acad. Sci. USA} \bibinfo{volume}{115},
  \bibinfo{pages}{8260--8265}.
%Type = Incollection
\bibitem[{Brandt et~al.(2007)Brandt, Ohtsuki, Iwasa and
  Sigmund}]{takeuchi2007mathematics}
\bibinfo{author}{Brandt, H.}, \bibinfo{author}{Ohtsuki, H.},
  \bibinfo{author}{Iwasa, Y.}, \bibinfo{author}{Sigmund, K.},
  \bibinfo{year}{2007}.
\newblock \bibinfo{title}{A survey of indirect reciprocity}, in:
  \bibinfo{editor}{Takeuchi, Y.}, \bibinfo{editor}{Iwasa, Y.},
  \bibinfo{editor}{Sato, K.} (Eds.), \bibinfo{booktitle}{Mathematics for
  ecology and environmental sciences}. \bibinfo{publisher}{Springer},
  \bibinfo{address}{Berlin}, p.~\bibinfo{pages}{30}.
%Type = Article
\bibitem[{Hilbe et~al.(2018)Hilbe, Schmid, Tkadlec, Chatterjee and
  Nowak}]{hilbe2018indirect}
\bibinfo{author}{Hilbe, C.}, \bibinfo{author}{Schmid, L.},
  \bibinfo{author}{Tkadlec, J.}, \bibinfo{author}{Chatterjee, K.},
  \bibinfo{author}{Nowak, M.A.}, \bibinfo{year}{2018}.
\newblock \bibinfo{title}{Indirect reciprocity with private, noisy, and
  incomplete information}.
\newblock \bibinfo{journal}{Proc. Natl. Acad. Sci. USA} \bibinfo{volume}{115},
  \bibinfo{pages}{12241--12246}.
%Type = Article
\bibitem[{Krellner et~al.(2021)}]{krellner2022pleasing}
\bibinfo{author}{Krellner, M.}, et~al., \bibinfo{year}{2021}.
\newblock \bibinfo{title}{Pleasing enhances indirect reciprocity-based
  cooperation under private assessment}.
\newblock \bibinfo{journal}{Artif. Life} \bibinfo{volume}{27},
  \bibinfo{pages}{246--276}.
%Type = Article
\bibitem[{Lee et~al.(2021)Lee, Murase and Baek}]{lee2021local}
\bibinfo{author}{Lee, S.}, \bibinfo{author}{Murase, Y.}, \bibinfo{author}{Baek,
  S.K.}, \bibinfo{year}{2021}.
\newblock \bibinfo{title}{Local stability of cooperation in a continuous model
  of indirect reciprocity}.
\newblock \bibinfo{journal}{Sci. Rep.} \bibinfo{volume}{11},
  \bibinfo{pages}{14225}.
%Type = Article
\bibitem[{Leimar and Hammerstein(2001)}]{leimar2001evolution}
\bibinfo{author}{Leimar, O.}, \bibinfo{author}{Hammerstein, P.},
  \bibinfo{year}{2001}.
\newblock \bibinfo{title}{Evolution of cooperation through indirect
  reciprocity}.
\newblock \bibinfo{journal}{Proc. R. Soc. B} \bibinfo{volume}{268},
  \bibinfo{pages}{745--753}.
%Type = Book
\bibitem[{Mackie et~al.(2014)Mackie, Moneti, Denny and
  Shakya}]{mackie2015social}
\bibinfo{author}{Mackie, G.}, \bibinfo{author}{Moneti, F.},
  \bibinfo{author}{Denny, E.}, \bibinfo{author}{Shakya, H.},
  \bibinfo{year}{2014}.
\newblock \bibinfo{title}{What are Social Norms? How are They Measured?}
\newblock \bibinfo{publisher}{UNICEF/UCSD Center on Global Justice Project
  Cooperation Agreement Working Paper}, \bibinfo{address}{San Diego, CA}.
%Type = Article
\bibitem[{Murase et~al.(2022)Murase, Kim and Baek}]{murase2022social}
\bibinfo{author}{Murase, Y.}, \bibinfo{author}{Kim, M.}, \bibinfo{author}{Baek,
  S.K.}, \bibinfo{year}{2022}.
\newblock \bibinfo{title}{Social norms in indirect reciprocity with ternary
  reputations}.
\newblock \bibinfo{journal}{Sci. Rep.} \bibinfo{volume}{12},
  \bibinfo{pages}{455}.
%Type = Article
\bibitem[{Nowak(2006)}]{nowak2006five}
\bibinfo{author}{Nowak, M.A.}, \bibinfo{year}{2006}.
\newblock \bibinfo{title}{Five rules for the evolution of cooperation}.
\newblock \bibinfo{journal}{Science} \bibinfo{volume}{314},
  \bibinfo{pages}{1560--1563}.
%Type = Article
\bibitem[{Nowak and Sigmund(1998)}]{nowak1998evolution}
\bibinfo{author}{Nowak, M.A.}, \bibinfo{author}{Sigmund, K.},
  \bibinfo{year}{1998}.
\newblock \bibinfo{title}{Evolution of indirect reciprocity by image scoring}.
\newblock \bibinfo{journal}{Nature} \bibinfo{volume}{393},
  \bibinfo{pages}{573--577}.
%Type = Article
\bibitem[{Nowak and Sigmund(2005)}]{nowak2005evolution}
\bibinfo{author}{Nowak, M.A.}, \bibinfo{author}{Sigmund, K.},
  \bibinfo{year}{2005}.
\newblock \bibinfo{title}{Evolution of indirect reciprocity}.
\newblock \bibinfo{journal}{Nature} \bibinfo{volume}{437},
  \bibinfo{pages}{1291--1298}.
%Type = Article
\bibitem[{Ohtsuki and Iwasa(2004)}]{ohtsuki2004should}
\bibinfo{author}{Ohtsuki, H.}, \bibinfo{author}{Iwasa, Y.},
  \bibinfo{year}{2004}.
\newblock \bibinfo{title}{How should we define goodness? -- reputation dynamics
  in indirect reciprocity}.
\newblock \bibinfo{journal}{J. Theor. Biol.} \bibinfo{volume}{231},
  \bibinfo{pages}{107--120}.
%Type = Article
\bibitem[{Ohtsuki and Iwasa(2006)}]{ohtsuki2006leading}
\bibinfo{author}{Ohtsuki, H.}, \bibinfo{author}{Iwasa, Y.},
  \bibinfo{year}{2006}.
\newblock \bibinfo{title}{The leading eight: social norms that can maintain
  cooperation by indirect reciprocity}.
\newblock \bibinfo{journal}{J. Theor. Biol.} \bibinfo{volume}{239},
  \bibinfo{pages}{435--444}.
%Type = Article
\bibitem[{Oishi et~al.(2021)Oishi, Miyano, Kaski and
  Shimada}]{oishi2021balanced}
\bibinfo{author}{Oishi, K.}, \bibinfo{author}{Miyano, S.},
  \bibinfo{author}{Kaski, K.}, \bibinfo{author}{Shimada, T.},
  \bibinfo{year}{2021}.
\newblock \bibinfo{title}{Balanced-imbalanced transitions in indirect
  reciprocity dynamics on networks}.
\newblock \bibinfo{journal}{Phys. Rev. E} \bibinfo{volume}{104},
  \bibinfo{pages}{024310}.
%Type = Article
\bibitem[{Okada(2020)}]{okada2020two}
\bibinfo{author}{Okada, I.}, \bibinfo{year}{2020}.
\newblock \bibinfo{title}{Two ways to overcome the three social dilemmas of
  indirect reciprocity}.
\newblock \bibinfo{journal}{Sci. Rep.} \bibinfo{volume}{10},
  \bibinfo{pages}{16799}.
%Type = Article
\bibitem[{Okada et~al.(2017)Okada, Sasaki and Nakai}]{okada2017tolerant}
\bibinfo{author}{Okada, I.}, \bibinfo{author}{Sasaki, T.},
  \bibinfo{author}{Nakai, Y.}, \bibinfo{year}{2017}.
\newblock \bibinfo{title}{Tolerant indirect reciprocity can boost social
  welfare through solidarity with unconditional cooperators in private
  monitoring}.
\newblock \bibinfo{journal}{Sci. Rep.} \bibinfo{volume}{7},
  \bibinfo{pages}{9737}.
%Type = Article
\bibitem[{Okada et~al.(2018)Okada, Sasaki and Nakai}]{okada2018solution}
\bibinfo{author}{Okada, I.}, \bibinfo{author}{Sasaki, T.},
  \bibinfo{author}{Nakai, Y.}, \bibinfo{year}{2018}.
\newblock \bibinfo{title}{A solution for private assessment in indirect
  reciprocity using solitary observation}.
\newblock \bibinfo{journal}{J. Theor. Biol.} \bibinfo{volume}{455},
  \bibinfo{pages}{7--15}.
%Type = Article
\bibitem[{Olejarz et~al.(2015)Olejarz, Ghang and Nowak}]{olejarz2015indirect}
\bibinfo{author}{Olejarz, J.}, \bibinfo{author}{Ghang, W.},
  \bibinfo{author}{Nowak, M.}, \bibinfo{year}{2015}.
\newblock \bibinfo{title}{Indirect reciprocity with optional interactions and
  private information}.
\newblock \bibinfo{journal}{Games} \bibinfo{volume}{6},
  \bibinfo{pages}{438--457}.
%Type = Article
\bibitem[{Perret et~al.(2021)Perret, Krellner and Han}]{perret2021evolution}
\bibinfo{author}{Perret, C.}, \bibinfo{author}{Krellner, M.},
  \bibinfo{author}{Han, T.A.}, \bibinfo{year}{2021}.
\newblock \bibinfo{title}{The evolution of moral rules in a model of indirect
  reciprocity with private assessment}.
\newblock \bibinfo{journal}{Sci. Rep.} \bibinfo{volume}{11},
  \bibinfo{pages}{23581}.
%Type = Article
\bibitem[{Quan et~al.(2022)Quan, Nie, Chen and Wang}]{quan2022keeping}
\bibinfo{author}{Quan, J.}, \bibinfo{author}{Nie, J.}, \bibinfo{author}{Chen,
  W.}, \bibinfo{author}{Wang, X.}, \bibinfo{year}{2022}.
\newblock \bibinfo{title}{Keeping or reversing social norms promote cooperation
  by enhancing indirect reciprocity}.
\newblock \bibinfo{journal}{Chaos Solit. Fractals} \bibinfo{volume}{158},
  \bibinfo{pages}{111986}.
%Type = Article
\bibitem[{Quan et~al.(2019)Quan, Yang, Wang, Yang, Wu and
  Dai}]{quan2019withhold}
\bibinfo{author}{Quan, J.}, \bibinfo{author}{Yang, X.}, \bibinfo{author}{Wang,
  X.}, \bibinfo{author}{Yang, J.B.}, \bibinfo{author}{Wu, K.},
  \bibinfo{author}{Dai, Z.}, \bibinfo{year}{2019}.
\newblock \bibinfo{title}{Withhold-judgment and punishment promote cooperation
  in indirect reciprocity under incomplete information}.
\newblock \bibinfo{journal}{EPL} \bibinfo{volume}{128}, \bibinfo{pages}{28001}.
%Type = Article
\bibitem[{Radzvilavicius et~al.(2021)Radzvilavicius, Kessinger and
  Plotkin}]{radzvilavicius2021adherence}
\bibinfo{author}{Radzvilavicius, A.L.}, \bibinfo{author}{Kessinger, T.A.},
  \bibinfo{author}{Plotkin, J.B.}, \bibinfo{year}{2021}.
\newblock \bibinfo{title}{Adherence to public institutions that foster
  cooperation}.
\newblock \bibinfo{journal}{Nat. Commun.} \bibinfo{volume}{12},
  \bibinfo{pages}{3567}.
%Type = Article
\bibitem[{Silver and Shaw(2018)}]{silver2018pint}
\bibinfo{author}{Silver, I.M.}, \bibinfo{author}{Shaw, A.},
  \bibinfo{year}{2018}.
\newblock \bibinfo{title}{Pint-sized public relations: The development of
  reputation management}.
\newblock \bibinfo{journal}{Trends Cogn. Sci.} \bibinfo{volume}{22},
  \bibinfo{pages}{277--279}.
%Type = Book
\bibitem[{Sugden(1986)}]{sugden1986economics}
\bibinfo{author}{Sugden, R.}, \bibinfo{year}{1986}.
\newblock \bibinfo{title}{The Economics of Rights, Cooperation and Welfare}.
\newblock \bibinfo{publisher}{Blackwell}, \bibinfo{address}{Oxford}.
%Type = Article
\bibitem[{Uchida(2010)}]{uchida2010effect}
\bibinfo{author}{Uchida, S.}, \bibinfo{year}{2010}.
\newblock \bibinfo{title}{Effect of private information on indirect
  reciprocity}.
\newblock \bibinfo{journal}{Phys. Rev. E} \bibinfo{volume}{82},
  \bibinfo{pages}{036111}.
%Type = Article
\bibitem[{Uchida and Sasaki(2013)}]{uchida2013effect}
\bibinfo{author}{Uchida, S.}, \bibinfo{author}{Sasaki, T.},
  \bibinfo{year}{2013}.
\newblock \bibinfo{title}{Effect of assessment error and private information on
  stern-judging in indirect reciprocity}.
\newblock \bibinfo{journal}{Chaos Solit. Fractals} \bibinfo{volume}{56},
  \bibinfo{pages}{175--180}.

\end{thebibliography}

\end{document}